\documentclass[aps,pra,notitlepage,nofootinbib,superscriptaddress,reprint]{revtex4-2}
\usepackage{graphicx}
\usepackage{multirow}
\usepackage[tbtags]{amsmath}
\usepackage{amssymb,amsthm,mathtools}
\usepackage{braket}
\usepackage{epstopdf,cancel}
\usepackage[normalem]{ulem}
\usepackage{epsf,latexsym,bbm,euscript}
\usepackage[colorlinks,bookmarks=false]{hyperref}

\usepackage{comment} 
\usepackage{xcolor} 
\usepackage{cancel}

\begin{document}

\author{Carolyn E. Wood}
\affiliation{Australian Research Council Centre for Engineered Quantum Systems, School of Mathematics and Physics, The University of Queensland, St Lucia, QLD 4072, Australia}
\author{Harshit Verma}
\affiliation{Eviden, Blk 988 Toa Payoh North, \#08-01, Singapore 319002}
\author{Fabio Costa}
\affiliation{Australian Research Council Centre for Engineered Quantum Systems, School of Mathematics and Physics, The University of Queensland, St Lucia, QLD 4072, Australia}
\affiliation{Nordita, Stockholm University and KTH Royal Institute of Technology, Stockholm, 106 91, Sweden}
\author{Magdalena Zych}
\affiliation{Australian Research Council Centre for Engineered Quantum Systems, School of Mathematics and Physics, The University of Queensland, St Lucia, QLD 4072, Australia}
\affiliation{Department of Physics, Stockholm University, AlbaNova University Center, SE-106 91 Stockholm, Sweden}

\title{Operational models of temperature superpositions}

\begin{abstract}
An interacting quantum system and thermal bath can reach thermal equilibrium, resulting in the system and bath acquiring the same temperature.
But how does a delocalised quantum system thermalise with a bath whose local temperature varies as, for example, in the Tolman effect? Here we formulate two scenarios in which the notion of a ``superposition of temperatures" may arise. First: a probe interaction with two different baths dependent on the state of another quantum system (a control). Second: a probe interaction with a single bath whose purified state is a superposition of states which each correspond to different temperatures. We show that the two scenarios are fundamentally different and can be operationally distinguished. Moreover, we show that the final probe state is sensitive to the specific realisation of the thermalising channels, and that our results hold for partial and pre-thermalisation cases. Our models may be applied to scenarios involving joint quantum, gravitational, and thermodynamic phenomena, and explain some recent results found in quantum interference of relativistic probes thermalising with Unruh or Hawking radiation. 
\end{abstract}
\maketitle

\section{Introduction}\label{sec:intro}
Temperature is well understood in a classical sense as a macroscopic property of the average kinetic energy of an ensemble of particles, while in quantum physics it is a key parameter defining the Gibbs state of a system---inextricably linked to the dynamics of thermalisation, and the operation of the laws of thermodynamics in that regime~\cite{VinjanampathyAnders2016}. 
Recent research into temperature's role in quantum thermodynamics has mainly focused on sensing using quantum systems (i.e. thermometry)~\cite{Mehboudi2019}, or the complementarity relation between temperature and energy fluctuations~\cite{MillerAnders2018}, and even an argument for describing temperature as a quantum operator~\cite{Ghonge2018}. 

In quantum thermometry, interferometry~\cite{Stace2010}---or some initial quantum coherence of a single-qubit probe~\cite{Jevtic2015}---appears to improve probe sensitivity. But can temperature itself exhibit quantum characteristics such as superposition or quantum coherence? It is natural to consider whether a situation exists where a quantum system could be delocalised and thus interact with baths of different temperatures. Alternately, one could imagine a superposition of states which are each associated with a different temperature.

There are certain scenarios at the intersection of quantum theory and thermodynamics with relativity in which these abstract ideas do in fact become relevant.  
See, for example, the Tolman--Ehrenfest effect~\cite{Tolman1930, TolmanEhrenfest1930}, in which locally-measured temperatures within a gas in thermal equilibrium in a gravitational field vary with position due to gravitational redshift, or time dilation. Because a quantum system cannot be infinitely well-localised it is implied that, fundamentally, such a system acting as a thermometer will be coherently spread over baths with different local temperatures. 
Another such scenario arises for thermalisation of matter coupled to radiation, such as in the Unruh and Hawking effects. There, internal states of some system in non-inertial motion thermalise to a temperature proportional to the system's acceleration. Again, taking into account that classical trajectories are just an approximation, one fundamentally deals with a system which is coherently spread over trajectories with different local accelerations each leading to a different thermal state of the probe system~\cite{Foo2020, Barbado2020, fooPRR2021}.
These scenarios suggest a need for a sharper understanding of thermal processes involving quantum systems: those where temperature does not take a fixed, classical value and thermalisation can occur in superposition.

In this work, we outline a general approach by which to address this idea of quantum features of temperature, and identify two cases that describe how a ``superposition of temperatures" might arise. In other words, we consider situations where the degrees of freedom of a system which are responsible for thermalisation have some form of coherence. The first case is best visualised as a probe system (or ``thermometer'') prepared in a superposition of two amplitudes, and interacting with an independent thermal bath for each of the amplitudes. In the second scenario, a single bath is in a superposition of purified states, each of which yields a different thermal state when reduced to the bath alone. These models assume the usual idealised probe--bath dynamics driven by the zeroth law of thermodynamics, where an interaction between the two results in a thermal state of the probe at the same temperature as that of the bath. We use the framework of quantum channels to derive the outcomes of the processes in the two scenarios and discuss both the conditions under which thermalisation can or can not be achieved, and the sense in which a notion of ``superposition of temperatures'' might arise. We also study partial or pre-thermalisation---where a quantum system far from equilibrium interacts with some environment for a finite time. We find, crucially, that the key features of the behaviour of the probe and coherence between temperatures are reproduced in the pre-thermalisation regime, showing our results do not depend on full thermalisation of the probe. Finally, we find that our two scenarios will naturally arise in the relativistic cases just discussed, and that the single-bath case in particular resolves an outstanding problem for relativistic quantum systems in superpositions of accelerations. 

The paper is laid out as follows: In Section~\ref{sec:defineprob}, we set up the necessary background, addressing challenges in defining thermalisation and superpositions in this context. In Section~\ref{sec:OpModels} we consider how to produce appropriate models, then explore the two models described above: The first, in Section~\ref{sec:TwoBath}, involving a single probe in superposition interacting with two independent baths, the second, in Section~\ref{sec:OneBath}, a single probe interacting with a single bath which is in a superposition. 
We explore the partial or pre-thermalisation case in Section~\ref{sec:partialpre}, before briefly revisiting the relativistic scenarios just mentioned in Section~\ref{sec:Rel}. We conclude by discussing the implications of the results in Section~\ref{sec:discussion}.

\section{Preliminaries and Definitions}\label{sec:defineprob}
We assume a quantum system interacting with some environment, with temperature arising as a consequence of some thermalisation process between the two.
We can characterise the thermalisation of an open quantum system as the action of a CPTP (completely positive trace-preserving) map between the system and environment.
For cases where full thermalisation is reached, both the \emph{initial} state of the environment or thermal bath\footnote{Though not restricted in this work to a particular interaction or environment state, as thermalisation is our main focus we still refer to the environment hereafter as a ``bath''.} (subscript B) and the \emph{final} state of the probe system (subscript S) are Gibbs states with the same temperature $T$~\cite{Lenard1978},~i.e.,
\begin{equation}\label{eq:Gibbs}
\rho^{\beta}_{B(S)} = \frac{e^{-\beta \hat{H}_{B(S)}}}{\mathcal{Z}^{\beta}},
\end{equation}
where $\beta = 1/k_B T$ is the inverse temperature with $k_B$ the Boltzmann constant, and $\mathcal{Z}^{\beta} = \text{Tr}~e^{-\beta \hat H_{B(S)}}$ the normalising partition function. $\hat H_{B(S)}$ is the Hamiltonian of the bath (probe).

Consider an initial input product state, in this case between our probe (S) and bath (B): $\rho_{\mathrm{in}} = \rho_B \otimes \rho_S$. The general form of a quantum channel, e.g.~\cite{NielsenChuang2010}, following action of the CPTP map $\mathcal{E}(\cdot)$, results in the Kraus representation~\cite{KrausBook1983} of the post-interaction state of the probe:
\begin{equation}\label{eq:thermalchannel}
\mathcal{E}(\rho_{S}) = \text{Tr}_B \left\{U_{BS} \rho_{\mathrm{in}} U_{BS}^{\dagger}\right\} \equiv \sum_{k,l} c_l^{\beta} M_{kl} \rho_S M_{kl}^{\dagger},
\end{equation}
where we choose a basis in which $\rho_B$ is diagonal: $\rho^{\beta}_B = \sum_l c_l^{\beta} \ket{l}\bra{l}$, and the Kraus operators arising from some unitary interaction between the bath and probe, $U_{BS}$, read $M_{kl} = \bra{k}U_{BS}\ket{l}$. 

For a thermalising channel specifically, the probe-bath interaction represented by $U_{BS}$ must leave the probe in a thermal state, e.g.~$\rho_S^{\beta}=\sum_{k} c^{\beta}_k \ket{k}\bra{k}$. A simple example would be a SWAP operator, $S_{BS}$, which `swaps' the states of the bath and probe system. For a probe initially in some arbitrary pure state $\rho_S = \ket{\eta}\bra{\eta}_S$, under the SWAP operator, $\sum_{k,l} c_l^{\beta} M_{kl} \rho_S M_{kl}^{\dagger} = \rho^{\beta}_S$, which is a thermal state on the probe at inverse temperature $\beta$.

We want to, however, consider a more general dilation of a thermalisation channel because an overall channel under quantum control is sensitive to the choice of dilation~\cite{Oi2003, Abbott2020}. Recall that Stinespring's dilation theorem~\cite{Stinespring1955} states that all dilations can be expressed as unitaries acting on an environment in a pure state, and that all such dilations are related to each other by isometries on the environment. Therefore, to construct a general dilation, we need to purify the bath by entangling it with some ancillary system (denoted A hereafter) to create an environment that is in a pure state~\cite{NielsenChuang2010}. 
These purified states are also known as thermofield double states in the context of finite-temperature quantum field theory~\cite{TakahashiUmezawa1975}\footnote{Also frequently used to represent the decomposition of a global state of a relativistic quantum field into local modes in the presence of a horizon~\cite{Israel1976, Crispino2008}.}. 
We write a purification of a specifically thermal state with inverse temperature $\beta$, in the energy eigenbasis, as
\begin{equation}
|\theta^{\beta}\rangle=\sum_{n} \sqrt{c_n^\beta}\left|n,n\right\rangle,
\label{eq:psitfd}
\end{equation}
where $c_n^\beta:=\frac{e^{-E_{n}\beta }}{\mathcal{Z}^\beta}$, and $E_n$ (for $n=0,1,...$) is the spectrum of the bath Hamiltonian. 
The state of the bath alone can be recovered by tracing out the ancillary system: $\rho_{B} := \mathrm{Tr}_{A}\{\ket{\theta^{\beta}}\bra{\theta^{\beta}}_{AB}\}$, where $\ket{\theta^{\beta}}_{AB}$ is the joint pure state of the bath and ancilla. 

The general dilation of the channel is then given by 
\begin{equation}\label{eq:primegenDilKraus}
\mathcal{E}'(\rho_S) = \text{Tr}_{AB}\Big\{U'_{A'B'S}\rho_{\text{in}}U'^{\dagger}_{A'B'S}\Big\},
\end{equation}
where primes indicate a larger Hilbert space such that $\dim(\mathcal{H}_{A' B'}) \geq \dim(\mathcal{H}_{AB})$, $\rho_{\text{in}} = \ket{\theta^{\beta}}\bra{\theta^{\beta}}_{AB} \otimes \rho_S$, and
\begin{equation}\label{eq:singlechannelU}
U'_{A'B'S}=(w \otimes\mathbb{I}_S)\cdot(\mathbb{I}_A\otimes U_{BS}).
\end{equation}

The arbitrary isometry $w$ is defined $w: \mathcal{H}_{AB}\rightarrow \mathcal{H}_{A'B'}$.

Hence Eq.~\eqref{eq:primegenDilKraus}, dropping the identities in Eq.~\eqref{eq:singlechannelU}, will be:
\begin{equation}\label{eq:genDilKraus}
\mathcal{E}'(\rho_S) = \text{Tr}_{A'B'}\Big\{w~ U_{BS}\ket{\theta}\bra{\theta}_{AB}\rho_{S} U^{\dagger}_{BS} w^{\dagger}\Big\}.
\end{equation}

Cycling the isometries under the trace allows for the substitution $w^{\dagger}w =\mathbb{I}$.
Hence, under the aforementioned SWAP operator and again with $\rho_S = \ket{\eta}\bra{\eta}$, we arrive at a final expression for the probe state:
\begin{equation}
\mathcal{E}'(\rho_S) =  \text{Tr}_{A'B'}\Big\{\ket{\theta^{\beta}}_{AS} \ket{\eta}_{B}\bra{\eta}_{B} \bra{\theta^{\beta}}_{AS} \Big\} = \rho^{\beta}_S,
\end{equation}
which is a thermal state which is not dependent on the particular dilation of the channel (i.e., the isometries), as expected for a single channel.

\section{Operational models of temperature superpositions}\label{sec:OpModels}
Ultimately, we are describing in this work a generalisation of thermalisation processes, motivated by the context of creating temperature superpositions. Of course, a system with some defined temperature is implied to be in a Gibbs state, whereas generally the notion of superposition only applies to two (or more) distinct pure states, for example, some $\ket{\psi}\neq \ket{\phi}$, placed in a linear combination, $\ket{\chi} = \alpha\ket{\psi} + \beta \ket{\phi}$ (where $\alpha$ and $\beta$ are non-vanishing complex numbers). 
However, operationally, we can think of the preparation of a superposition as a process where a state of one or more different DoFs is created in response to a quantum state of some other particular DoF, which we will later refer to as a control. By preparing and measuring the control in superposition, the other DoF(s) are prepared in the corresponding superposition. 

Therefore, one operational context in which the notion of ``superposition of temperatures'' arises is when one has a superposition of purifications such as $\ket{\theta^{\text{sup}}} \propto \ket{\theta^0}+\ket{\theta^1}$, where, for $x=0,1$, the state $\ket{\theta^x}$ is a purification of a thermal state at a temperature $T_x$. 

A second context is that of quantum controlled channels, e.g.~\cite{Oi2003}, where in our case here we have a system interacting with one of two baths each at a different temperature, depending on the state of some control DoF. We look at this scenario first.

\subsection{Two baths: Superpositions of thermal channels}\label{sec:TwoBath}

For our first case of a superposition of temperatures we define an interaction between a probe system and a composite purified bath $\rho_{AB}$ comprised of two subsystems in a product state $\rho_{AB} = \rho_{A_0B_0} \otimes \rho_{A_1B_1}$. This can arise, for example, if the subsystems are thermally isolated from each other, at least over the time-scale of the interaction with the probe. This model would be relevant to a quantum system probing a Tolman-affected bath, as we discuss further in Section~\ref{sec:Rel}.

This is called quantum-controlled thermalisation because a quantised control DoF determines which subsystem the probe thermalises with. The probe system (S), baths (AB), and control (C) thus form the initial product state $\rho_{\text{in}} = \rho_{AB} \otimes \rho_C \otimes \rho_S$. 
A unitary operator $U'_{A'B'CS}$ for the entire process, by Eq.~\eqref{eq:singlechannelU}, should have the form $U'_{A'_0B'_0S}\otimes\ket{0}\bra{0}_C+U'_{A'_1B'_1S}\otimes\ket{1}\bra{1}_C$, where the control states $\ket{i}_C$, $i=0,1$ are orthonormal and determine that the probe interacts with the corresponding bath $B_i$ through unitary $U_{B_iS}$.

\begin{figure}
  \includegraphics[width=0.44\textwidth]{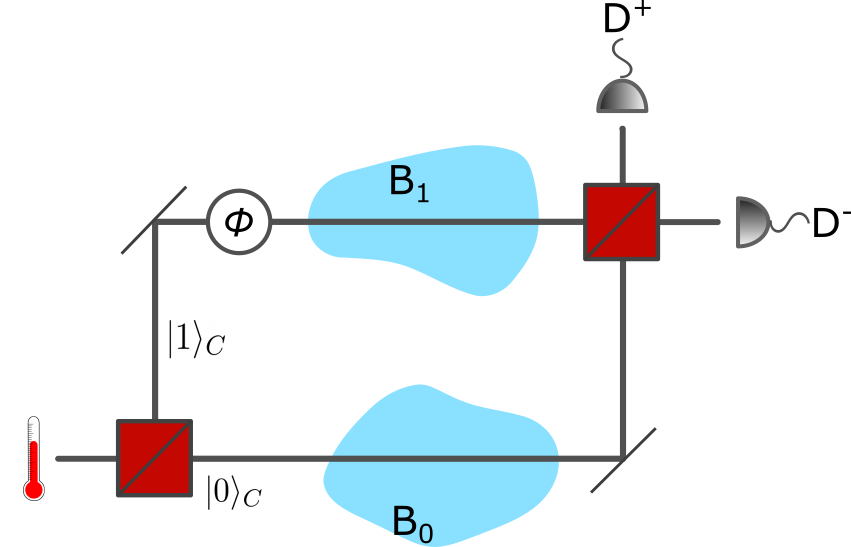}
  \caption{A Mach-Zehnder interferometer as an example realisation of the two-bath case. An input probe state enters from the left, is placed in a superposition, and travels along the two arms of the interferometer, on each of which is a bath thermalised to some temperature. The paths recombine at the second beamsplitter, and the output state is then detected at $D^+, D^-$. \label{fig:2BathMZInterferometer}}
\end{figure}

While this is independent of any specific framework, the MZ interferometer of Figure~\ref{fig:2BathMZInterferometer} is certainly a potential realisation of the scenario. In that example, the two states $\ket i_C$ of the control are identified with the two paths through the interferometer, the probe is an internal DoF of the interfering system (which could be, for example, a particle) and it thermalises with subsystems $B_0$ or $B_1$ depending on the path taken.

Our $U'_{A'B'CS}$ will include the $w$ isometry as in Eq.~\eqref{eq:singlechannelU}, but now that we have two subspaces, associated with baths $0$ and $1$, we must introduce an additional isometry, $v: \mathcal{H}_{AB}\rightarrow \mathcal{H}_{A'B'}$. This ensures that all elements of the channel are embedded in the same Hilbert space. 

The unitary operator for the entire process is then:
\begin{multline}\label{eq:fullU}
U'_{A'B'CS} = (w_{0}\otimes v_{1}\otimes \mathbb{I}_S) (\mathbb{I}_{A_0} \otimes U_{B_0S} \otimes \mathbb{I}_{A_1B_1}) \otimes \ket{0}\bra{0}_C\\ + (w_{1}\otimes v_{0}\otimes \mathbb{I}_S) (\mathbb{I}_{A_1} \otimes U_{B_1S}\otimes \mathbb{I}_{A_0B_0}) \otimes \ket{1}\bra{1}_C,
\end{multline}
where the $i=0,1$ subscripts on both isometries identify them with their respective channels.

We take the initial state of the control to be $\rho_C=\ket+\bra+$, where $\ket+=\frac{1}{\sqrt{2}}(\ket0+\ket1)$, and measure it in the superposition $\ket{\phi} = \frac{1}{\sqrt{2}}\left(\ket{0} + e^{i\phi}\ket{1}\right)$, with relative phase $\phi$ assumed to be fully controllable. 

The sub-normalised output state of the probe is then $\rho_S{(\phi)}= \text{Tr}_{A'B'} \left\lbrace \bra{\phi}_C U' \rho_{\mathrm{in}}(0) U'^{\dagger} \ket{\phi}_C \right\rbrace $, which yields four terms:
\begin{multline}\label{eq:qctrlchan}
\rho_S{(\phi)}= \frac{1}{4}\text{Tr}_{A'B'} \Big\lbrace (w_0U_{B_0S}\otimes v_1)\rho_{ABS} (U^{\dagger}_{B_0S}w^{\dagger}_0 \otimes v^{\dagger}_1) \\
+ (w_1U_{B_1S}\otimes v_0)\rho_{ABS} (U^{\dagger}_{B_1S}w^{\dagger}_1 \otimes v^{\dagger}_0) \\
+ \Big(e^{i\phi} (w_0U_{B_0S}\otimes v_1)\rho_{ABS} (U^{\dagger}_{B_1S}w^{\dagger}_1 \otimes v^{\dagger}_0) + \text{H.c.}\Big) \Big\rbrace,
\end{multline}
and H.c. refers to the Hermitian conjugate.

The first two terms in Eq.~\eqref{eq:qctrlchan} are each a channel arising from the interaction $U_{B_iS}$ between the system and bath $B_i$, with the form of a single quantum channel as in Section~\ref{sec:defineprob}.
The other two terms are ``cross-terms'' between channels, with the third term in Eq.~\eqref{eq:qctrlchan}, for arbitrary pure system state $\rho_S = \ket{\eta}_S\bra{\eta}_S$:
\begin{multline}\label{eq:2baththirdterm}
\frac{1}{4}\text{Tr}_{A'B'}\Big\lbrace e^{i\phi}w_{0} U_{B_0S} \big( \ket{\theta^{\beta_0}}_{A_0B_0}\ket{\eta}_S \big) v_1 \ket{\theta^{\beta_1}}_{A_1B_1}\\ \bra{\theta^{\beta_0}}_{A_0B_0} v^{\dagger}_0 \big(\bra{\eta}_S \bra{\theta^{\beta_1}}_{A_1B_1}\big) U^{\dagger}_{B_1S} w^{\dagger}_1 \Big\rbrace.
\end{multline}
The fourth term is of course its Hermitian conjugate. 

As in Section~\ref{sec:defineprob} we take the swap operator $S_{B_iS}$, which simply swaps the state of the $i^{\mathrm{th}}$ bath with the state of the system. 
So the first two terms of Eq.~\eqref{eq:qctrlchan} are each thermal states on the probe $\rho^{\beta_i}_S$ at the respective temperatures of the baths, and Eq.~\eqref{eq:2baththirdterm} becomes:
\begin{multline}
\frac{1}{4} e^{i\phi} \bra{\theta^{\beta_0}}_{A_0B_0} v^{\dagger}_0 w_{0} \Big( \ket{\theta^{\beta_0}}_{A_0S}\ket{\eta}_{B_0} \Big)\\ \Big(\bra{\eta}_{B_1} \bra{\theta^{\beta_1}}_{A_1S}\Big)  w^{\dagger}_1 v_1\ket{\theta^{\beta_1}}_{A_1B_1},
\end{multline}
with the fourth term again its Hermitian conjugate.

The theta kets and bras form partial inner products, but this is as far as we can reduce these cross-terms without further specifying the involved systems.

Defining:
\begin{equation}\label{eq:BigPhi}
\ket{\Phi_i} := \bra{\theta^{\beta_i}}_{A_iB_i} v^{\dagger}_i w_i \Big(\ket{\theta^{\beta_i}}_{A_iS}\ket{\eta}_{B_i}\Big),
\end{equation}
we can write the final (sub-normalised) state of the probe as:
\begin{multline}\label{eq:2bathfinaliso}
\rho_S{(\phi)} = \frac{1}{4}\Big[\rho^{\beta_0}_S + \rho^{\beta_1}_S + \Big(e^{i\phi}\ket{\Phi_0}\bra{\Phi_1} + \text{H.c.}\Big)\Big].
\end{multline}
Note that Eq~\eqref{eq:BigPhi} is also sub-normalised, as $\braket{\Phi_i|\Phi_i} \leq 1$.

The first two terms of Eq.~\eqref{eq:2bathfinaliso} taken individually are thermal states at the temperature of each bath. The other two terms are not in general thermal, depending not just on the initial states of both baths and of the system, but also on the isometries associated with each channel dilation, as expected. It is also important to note that the isometries also ensure that we are free to choose any unitary to represent the probe-bath interaction, and that Eq.~\eqref{eq:2bathfinaliso} does not depend on the SWAP operation.
See Appendix~\ref{sec:appendix} for the special case where the isometries are unitaries on the bath, which is similar to the result of Ref.~\cite{Oi2003} for generic (i.e., not thermal) quantum channels in superposition.

One might have thought that if both baths are at the same temperature, i.e.~for $\beta_0 = \beta_1\equiv\beta$, the system simply thermalises to that temperature, however Eq.~\eqref{eq:2bathfinaliso} explicitly demonstrates that this is not the case unless $v^{\dagger}_iw_i = \mathbb{I}$, or we tune $\phi$ to the particular value that destroys the interference of the control. 

In cases where $v$ and $w$ do form an identity, we can further reduce the partial inner products of Eq.~\eqref{eq:BigPhi}, and arrive at the corresponding special case of Eq.~\eqref{eq:2bathfinaliso}:
\begin{multline}
\rho_S(\phi) = \frac{1}{4} \Big[ \rho^{\beta_0}_S + \rho^{\beta_1}_S + \Big(e^{i\phi}\rho_S^{\beta_0} \ket{\eta}_{B_0}\bra{\eta}_{B_1}\rho_S^{\beta_1} + \text{H.c.}\Big)\Big].
\end{multline}

So we see that the form of Eq.~\eqref{eq:2bathfinaliso} has crucial implications for the system's coherence, and thus the extent to which one can speak about any kind of coherence in the final state of the probe (or of the temperature). This result can be understood through the complementarity between quantum coherence and which-path information---via the analogy with the Mach-Zehnder interferometer in Fig.~\ref{fig:2BathMZInterferometer}. In an interferometric scenario like in Fig.~\ref{fig:2BathMZInterferometer} the visibility of the final interference pattern contains relevant information about the dynamics and correlations that develop between the involved systems. 
Recall that visibility is quantified by the magnitude of the off-diagonal elements of the final state~\cite{Mandel1991, Englert1996}. Here that final state is that of the control, and the visibility is equivalently the contrast of the interference pattern obtained by measuring the control in a superposition basis. 

The probability of measuring the control in the state $\ket\phi$ is $\text{P}{(\phi)} = \text{Tr}\{\rho_S{(\phi)}\}$; explicitly:
\begin{equation}\label{eq:2bathisoprob}
\text{P}{(\phi)} = \frac{1}{2} + \frac{1}{2 }\Big| \text{Tr}\Big\{\ket{\Phi_0}\bra{\Phi_1}\Big\} \Big| \cos(\phi + \psi),
\end{equation}
where $\psi$ is a phase defined via $\text{Tr}_S \{\ket{\Phi_0}\bra{\Phi_1}\}\equiv |\text{Tr}_S \{\ket{\Phi_0}\bra{\Phi_1}\}|e^{-i\psi}$.
 
The visibility, therefore, is 
\begin{equation}\label{eq:2bathvisibilityiso}
\mathcal{V} = \Big|\braket{\Phi_1|\Phi_0} \Big|.
\end{equation}

Crucially, this visibility depends not just on the temperatures of the two baths, but also the isometries $v$ and $w$:
\begin{multline}\label{eq:2bathvisexpanded}
\braket{\Phi_1|\Phi_0} = \bra{\theta^{\beta_0}}_{A_0B_0} v^{\dagger}_0 w_0 \big(\sqrt{\rho^{\beta_0}\rho^{\beta_1}} \otimes \ket{\eta}_{B_0}\bra{\eta}_{B_1}\big) \\ w^{\dagger}_1v_1\ket{\theta^{\beta_1}}_{A_1B_1}.
\end{multline}
Again, if $v^{\dagger}_i w_i=\mathbb{I}$, for $\beta^0 = \beta_1 \equiv \beta$ the above reduces further to $\bra{\eta}(\rho^{\beta})^2\ket{\eta} = \text{Tr} (\rho^{\beta})^2$.

Note that in general the visibility is less than 1. We know by the Cauchy--Schwarz inequality, $|\braket{\Phi_1|\Phi_0}| \leq ||\Phi_1||~||\Phi_0||$, that visibility can only equal $1$ if the two states are normalised and equal (up to a phase).

Taking Eq.~\eqref{eq:2bathvisexpanded}:
\begin{equation}\label{eq:2bathCauchyPhi}
||\Phi_0|| = \bra{\theta^{\beta_0}}_{A_0B_0} v^{\dagger}_0 w_0 \big( \rho^{\beta_0} \ket{\eta}\bra{\eta}_{B_0} \big) w^{\dagger}_0v_0\ket{\theta^{\beta_0}}_{A_0B_0},
\end{equation}
consider the general structure of this expression: $\bra{\psi}\rho\ket{\psi}$, for $\braket{\psi|\psi} \leq 1$. We know that since $\rho \geq 1$ and $\text{Tr}\rho = 1$, the eigenvalues of $\rho$ cannot be larger than one, so $\bra{\psi}\rho\ket{\psi}=1$ implies that $\rho = \ket{\psi}\bra{\psi}$ and $\braket{\psi|\psi}=1$. 

This implies the condition 
\begin{equation}
\rho^{\beta_0}\ket{\eta}\bra{\eta}_{B_0} = w^{\dagger}_0 v_0\ket{\theta^{\beta_0}}\bra{\theta^{\beta_0}}v^{\dagger}_0 w_0,
\end{equation}
and for this to hold the system and bath must both be in a pure state, which can only be achieved at zero temperature (and with a non-degenerate ground state). So, with $\ket{\theta^{\beta_0}} = \ket{0}\ket{0}$ signifying a state with these conditions, Eq.~\eqref{eq:2bathvisexpanded} becomes
\begin{equation}
\braket{\Phi_1|\Phi_0} = \bra{0}\bra{0} v^{\dagger}_0 w_0 \ket{0} \ket{\eta}_{B_0} \bra{0} \bra{\eta}_{B_1} w^{\dagger}_1v_1\ket{0}\ket{0}.
\end{equation}
This expression can only be equal to one when the combined isometries rotate the system to the vacuum state, i.e. $v^{\dagger}_i w_i \ket{0}\ket{\eta} = \ket{0}\ket{0}$.

Hence, in general, the visibility cannot reach unity, directly showing that the probe--bath(s) interactions do affect the coherence of the control even for equal bath temperatures. In other words, even for identical isometries and baths at the same temperature, the bath subsystems still store which-way information~\cite{Englert1996} about the control, except in the above special case of both baths at zero temperature.

The above also holds for the special case where the isometries are unitaries, as detailed in Appendix~\ref{sec:appendix}. In that case the visibility is
\begin{equation}\label{eq:2bathvisibilityunitariesmain}
\mathcal{V} = \left|\text{Tr} \left\lbrace u^{0\dagger}\rho_S^{\beta_0} \rho_S^{\beta_1} u^1 \rho_S \right\rbrace \right|,
\end{equation}
where $u^i$ are unitaries associated with each bath. 
The maximum visibility is 
\begin{equation*}
\mathcal{V}_{\mathrm{max}}^{\mathrm{pure}} = c^{\beta_0}_0 c^{\beta_1}_0,
\end{equation*}
for an initial pure state of the probe. The intuition is again that visibility is maximised when the local unitaries effectively rotate probe states $\ket{s}$ to the energy ground state (up to a phase) as it has the highest overlap with a thermal state at any finite temperature. Thus, for an arbitrary mixed state, the maximum visibility reads
\begin{equation}\label{eq:Vmax2bathmain}
\mathcal{V}_{max} =\sum_s p_s c^{\beta_0}_s c^{\beta_1}_s.
\end{equation}
Without loss of generality, the probabilities $p_s$ are ordered decreasingly: $p_s\geq p_{s+1}$. This is because thermal weights are likewise ordered decreasingly, i.e.~$c_k^{\beta_i}>c_{k+1}^{\beta_i}$, and due to the orthogonality of the states $\ket s$, a unitary $u^i$ can only rotate one of the states from an orthogonal set to the energy ground state. That being said, the same strategy can be iterated to find a unitary mapping of the eigenstates of $\rho_S$ to progressively higher energy eigenstates.

\subsection{One bath: A superposition of purifications}\label{sec:OneBath}

\begin{figure}
  \includegraphics[width=0.44\textwidth]{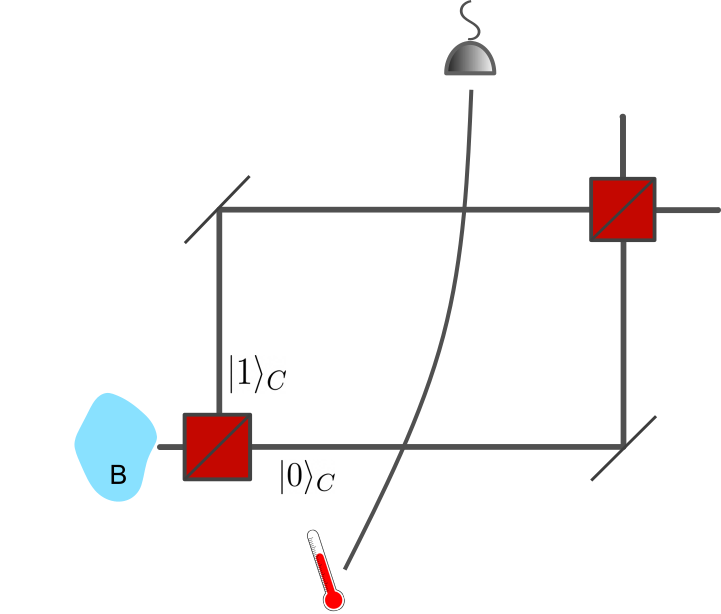}
  \caption{A diagram for the one-bath case in analogy to Figure 1. The bath is placed in a superposition and acquires a different temperature depending on the control state $\ket{0}_C$ or $\ket{1}_C$. The probe interacts/thermalises with the bath while the bath is in superposition. \label{fig:1BathFigure}}
\end{figure}

The previous case used quantum control of thermal channels associated with two separate bath subsystems as an operational model of a temperature superposition. Here, we explore the other model identified at the beginning of this section---where the bath is a single system in a superposition of purifications each corresponding to a different temperature. Figure~\ref{fig:1BathFigure} provides a sketch of such a situation. This scenario is highly relevant to accelerating quantum systems which experience the Unruh effect, as we discuss in Section~\ref{sec:Rel}. The procedure also has some similarity with the ``superposition of thermal states'' in optics, considered in ref.~\cite{Jeong2007}, however no superpositions of different temperatures were considered therein.

Purifications were defined in Section~\ref{sec:defineprob}, and Eq.~\eqref{eq:psitfd}. We now take a more general form
\begin{equation}\label{eq:singlepurification}
\ket{\theta^{\beta_x}(x)} = \frac{1}{\sqrt{2}} \sum_b e^{-i\phi_x} \sqrt{c_b^{\beta_x}} \ket{b,a(b,x)},
\end{equation}
where $x$ signifies that the purification gives rise to a thermal state at temperature $\beta_x$, and where $\{\ket{a(b,x)}\}_b$ is an orthonormal basis for the ancillary system. This basis can in principle be different for the different purifications $x = \{0,1\}$; phase $\phi_x \in \mathbb{R}$. 

An unnormalised superposition of Eq.~\eqref{eq:singlepurification} then reads $\ket{\psi} = \sum_{x=0,1} \ket{\theta^{\beta_x}(x)}$. One can also prepare the superposition of purifications using another DoF as a control, such that $\ket{\widetilde{\psi}} = \frac{1}{\sqrt{2}}\sum_{x=0,1} \ket{\theta^{\beta_x}(x)}\ket{x}_C$. For Eq.~\eqref{eq:singlepurification} that is
\begin{equation}\label{eq:suppurificationPhi}
\ket{\widetilde{\psi}} = \frac{1}{\sqrt{2}} \sum_{x=0,1} \sum_b e^{-i\phi_x} \sqrt{c_b^{\beta_x}} \ket{b,a(b,x)}\ket{x}_C.
\end{equation}

We could also choose to prepare an initial state for the bath by first measuring the control in the superposition $\ket{\phi}_C = \frac{1}{\sqrt{2}}\left(\ket{0}_C + e^{i\phi_C}\ket{1}_C\right)$ after tracing out the ancilla: $\rho_B^{\mathrm{in}}(\phi)= \bra{\phi}_C \text{Tr}_{A(x)}\{\ket{\widetilde{\psi}}\bra{\widetilde{\psi}}\}\ket{\phi}_C$. This would yield a reduced state of the bath such as:
\begin{multline}
\rho_B^{\mathrm{in}}(\phi) = \frac{1}{4} \Big[\rho^{\beta_{0}} + \rho^{\beta_{1}}\\
 + \Big(e^{-i\widetilde{\phi}} \sum_{b,b'} \sqrt{c_b^{\beta_{0}}c_{b'}^{\beta_{1}}} V^{0 1}_{b b'}\ket{b}\bra{b'} + \mathrm{H.c.}\Big)\Big],
\end{multline}
where $\widetilde{\phi} = \phi_0-\phi_1-\phi_C$, and we define $V^{x x'}_{bb'}:= \braket{a(b',x')|a(b,x)}$. $V^{xx'}_{bb'}$ is a matrix element of a special unitary operator $V^{x x'}$ transforming between ancilla bases associated with control states $x$ and $x'$. It satisfies $V^{x x}_{b b'} = V^{x' x'}_{b b'} = \delta_{b b'}$, and $V^{x x'} = (V^{x' x})^{\dagger}$.
If the purifications are the same for both ancilla states, $V^{xx'}$, for all $x,x'$, reduces to a delta function $\delta_{bb'}$. 

For the analysis here, however, we do not begin with the reduced state of the bath, instead using the Eq.~\eqref{eq:suppurificationPhi} joint state of bath, ancillary system, and control:
\begin{multline}
\rho_{ABC}^{\mathrm{in}} = \frac{1}{2} \sum_{x,x'= 0,1} \sum_{b,b'} e^{-i(\phi_{x}-\phi_{x'})}\sqrt{c_b^{\beta_{x}}c_{b'}^{\beta_{x'}}}\\ 
\ket{a(b,x)}\ket{b}\ket{x}\bra{x'}_C \bra{b'} \bra{a(b'x')},
\end{multline}
where the entire initial state is $\rho^{\mathrm{in}}=\rho_{ABC}\otimes \rho_S$. Again, we have a generic probe-bath unitary interaction $U_{BS}$. To generalise to arbitrary Kraus decompositions we simply introduce additional unitary operators $u^x_{AB}$ to the local interaction with the bath, which depend on the control and also act on both the bath and the ancillary system. 
Hence, the entire interaction, extended to include these ``local unitaries'', takes the form
\begin{equation}\label{eq:UBACS}
\widetilde{U}_{ABCS} = \sum_{x=0,1} \left[\left(u^{x}_{AB} \otimes \mathbb{I}_S \right) \cdot \left(\mathbb{I}_A\otimes U_{BS}\right)\right] \otimes \ket{x}\bra{x}_C.
\end{equation}

This $\widetilde{U}_{ABCS}$ is already an arbitrary dilation as the ancillae themselves can be of arbitrary dimensions, so we do not need to employ the isometries as in the two-bath scenario.

The final probe state, conditioned on the control $\ket{\phi}_C = \frac{1}{\sqrt{2}}\left(\ket{0}_C + e^{i\phi_C}\ket{1}_C\right)$, is
\begin{equation}
\widetilde{\rho}_S(\phi)=\bra{\phi}_C\text{Tr}_{AB} \left\{ \widetilde{U}_{ABCS} \rho^{\mathrm{in}}_{ABC}\otimes\rho_S \widetilde{U}_{ABCS}^{\dagger} \right\}\ket{\phi}_C.
\end{equation} 

Taking the SWAP operator once more to be our $U_{BS}$ 
this is, explicitly,
\begin{multline}
\widetilde{\rho}_S{(\phi)} = \frac{1}{4} \Big[\rho^{\beta_0}_S + \rho^{\beta_1}_S\\  
+ \Big(e^{-i\widetilde{\phi}} \sum_{b,b'} \sqrt{c_b^{\beta_{0}}c_{b'}^{\beta_{1}}} W^{10}_{bb'}\ket{b}\bra{b'} + \mathrm{H.c.} \Big)\Big],
\end{multline}
where $W^{xx'}_{bb'}:= \bra{a(b',x')}\text{Tr}_S\{u^{1\dagger}_{AS}u^0_{AS}\rho_S\}\ket{a(b,x)}$, similar to the $V^{x x}_{b b'}$ discussed above, but now post-interaction, and so dependent on the extra unitaries and the state of the probe.

We immediately see that, in this model too, the final probe state still depends on the temperatures of both baths, the initial state of the probe, and the interaction dynamics as represented by the local unitaries. Again the first two terms correspond to thermal states at each of the involved temperatures while the second two ``cross terms'' do not, however it is in these two cross terms that there are marked differences to the model of Section~\ref{sec:TwoBath}. 

Unlike in the two-bath case, here the probe \emph{can thermalise}. Thermalisation can occur when $\beta_0 = \beta_1$ and $W^{01}_{bb'}\propto\delta_{bb'}$. These criteria are met when: 1) both amplitudes of the purification (or superposed purifications) correspond to the same thermal state of the bath, 2) the purification basis is the same for both amplitudes, and 3) when $u^{1\dagger}_{AS}u^0_{AS} = e^{i\psi}\mathbb{I}$ (in which case the matrix $W^{0 1}_{b b'}$ is the identity). 

Looking at the visibility of the interference between the amplitudes of the control DoF, we find the probability to measure the control in $\ket{\phi}_C$ is
\begin{equation} \label{probtilde}
\tilde P{(\phi)} = \frac{1}{4} \Big[2 + \Big(e^{-i\widetilde{\phi}} \sum_b \sqrt{c_b^{\beta_{0}}c_{b}^{\beta_{1}}} W^{0 1}_{b b} + \mathrm{H.c.} \Big)\Big],
\end{equation}
which results in the visibility
\begin{equation}\label{eq:OneBathVis}
\widetilde{\mathcal{V}} = \left| \sum_b \sqrt{c_b^{\beta_0} c_b^{\beta_1}} W^{0 1}_{b b}\right|.
\end{equation}

There always exist $u_{AS}^x$ for which the visibility vanishes. Considering eigenstates $\ket s$ of $\rho_S$, it suffices to take $u_{AS}^i$ which map each $\ket s$ to different energy eigenstates of the probe, i.e.~such that $u_{AS}^1\ket s$ is orthogonal to $u_{AS}^0\ket s$ for every $s$. The same argument applies to the two-bath case under the conditions discussed in Section~\ref{sec:TwoBath} and Appendix~\ref{sec:appendix}.

Considering maximum visibility, on the other hand, firstly, it is easy to see that 
\begin{equation}
\Big|W^{01}_{bb'}\Big|= \Big|\bra{a(b',x')}\mathrm{Tr}_S\{ u_{AS}^{0\dagger} u_{AS}^1 {\rho_S}\}\ket{a(b,x)} \Big| \leq 1,
\end{equation}
where the inequality is saturated---thus maximising the visibility---when  $u_{AS}^{0\dagger} u_{AS}^1 = e^{i\varphi} \mathbb{I}$, for arbitrary $\varphi \in \mathbb{R}$. 
This means the maximum visibility is, therefore,
\begin{equation} \label{eq:onebathmaxvis}
\widetilde{\mathcal{V}}_{\mathrm{max}} = \sum_b \sqrt{c_b^{\beta_0}c_b^{\beta_1}}.
\end{equation}

Compare this with the analogous scenario in the two-bath case where the isometries are assumed to be unitaries, as discussed in the previous section, and in Appendix~\ref{sec:appendix}, where the maximum visibility, Equation~\eqref{eq:Vmax2bathmain}, is $\mathcal{V}_{\mathrm{max}} = \sum_s p_s c^{\beta_0}_s c^{\beta_1}_s$. The terms $p_s$ are probabilities arising from the diagonal form of the density matrix representation of the initial probe state. 
This is markedly different to the above one-bath case as it explicitly depends on the initial state of the probe.

Interestingly, Eq.~\eqref{eq:onebathmaxvis} coincides with the fidelity between the two thermal states, $\widetilde{\mathcal{V}}_{\mathrm{max}} = \textrm{Tr} \sqrt{\sqrt{\rho^{\beta_0}} \rho^{\beta_1} \sqrt{\rho^{\beta_0}}} = \textrm{Tr} \sqrt{\rho^{\beta_0} \rho^{\beta_1}}$, where the latter simplification is possible because the two states commute \cite{Heinosaari2011}.

When $\beta_0 =\beta_1=\beta$, independent of the actual value of $\beta$, $\widetilde{V}_{\mathrm{max}}=1$. However, even when $\beta_0=\beta_1$, the visibility in Eq.~\eqref{eq:OneBathVis} will be less than 1 when $W^{01}_{bb}\neq 1$, that is, when the purification  bases are not the same.

A further particular case of interest is probe-bath interactions which are independent of the control. This renders $u^{x}_{AB}$  independent of $x$, and the visibility is simply: $\tilde{\mathcal{V}}=\sum_b \sqrt{c_b^{\beta_0} c_b^{\beta_1}}\big|\braket{a(b,1)|a(b,0)}\big|$. This corresponds to a situation where the probe interacts with the bath after it has been prepared in a superposition, that is, after a particular state of the control has been detected post interference. 

In summary, we generally find that in the two-bath scenario, full thermalisation in general cannot be attained, while in this single-bath scenario it can, with particular conditions discussed above.

\section{Partial thermalisation}\label{sec:partialpre}

The two scenarios considered in the previous section assumed full thermalisation, which can be seen as the asymptotic future of any physical process which would in practise take finite time. In this section we look at partial, or pre-, thermalisation processes corresponding to our two models, demonstrating that the general features identified in the previous section also hold for finite interaction times.

To model the process of pre-thermalisation through a unitary interaction between the bath and the probe, we turn to the theory of thermal attenuators (TA)~\cite{Rosati2018}.  
In general, TAs represent the effect of a (thermal) lossy environment acting on a probe. In the most generic case, this environment could be constituted by bosonic DoFs in a thermal state. However, for studying pre-thermalisation it will be sufficient for us to consider a qubit probe interacting with an environment also constituted by qubit(s) (we refer to these as ``bath qubits'' through this section). In this case, a TA channel is a generalized amplitude damping channel (GADC) on the probe. 

The unitary interaction between the probe and bath which yields a GADC is of the form:
\begin{align}\label{eq:GADC}
U^{\eta}_{BS}=\left(\begin{array}{cccc}
1 & 0 & 0 & 0 \\
0 & \sqrt{1-\eta} & \sqrt{\eta} & 0 \\
0 & -\sqrt{\eta} & \sqrt{1-\eta} & 0 \\
0 & 0 & 0 & 1
\end{array}\right)~,
\end{align}
where $\eta$ parametrises the strength of the interaction. The action of a GADC on a probe can also be expressed in the form of Kraus operators, per Eq.~\eqref{eq:thermalchannel}. We note that in that case the purified bath state will consist of two qubits (one of them being the ancilla for purification). Such a purified state of the bath, constituted by qubits, has the form of Eq.~\eqref{eq:psitfd} with $n={0,1}$ and $\mathcal{Z}^\beta=e^{-E_0\beta}+e^{-E_1\beta}$.
The unitary interaction between the (purified) bath and the probe is then given as follows:
\begin{equation}
U_{BS}=U^{\eta}_{BS}\otimes \mathbb{I}_{A}~,
\label{eq:USB}
\end{equation}
where $U_{BS}^\eta$ acts on the probe and the bath qubit and the full interaction is an identity on the purification ancilla.

Hence, we model the process of thermalisation of a finite dimensional quantum system by the application of a GADC. If the interaction parameter is set to be the maximum ($\eta=1$), the probe thermalises after a single interaction with the bath, or equivalently with a single application of the GADC, thus here we set $\eta<1$. Therefore, after a single interaction with the bath, the probe acquires an intermediate state between the initial state and a thermal one.

To model the gradual approach to full thermalisation, we further consider a collisional model \cite{Scarani2002}, where our probe systems interact through the above mentioned GDAC with independent bath subsystems in the initial state $|\theta^\beta\rangle$. Specifically, the probe interacts with each of these subsystems individually and successively through the unitary operator $U_{BS}$. Notably, this corresponds to multiple consecutive applications of the GADC on the probe such that each application represents a time step. We generally expect that after a sufficient number of such steps, the probe assumes a thermal state and continues to remain in that state even if the process is continued. Indeed, this does hold, as we show in the following sections.

\subsection{Collisional model with generalized amplitude damping channel (GADC)}
\begin{figure}[t]
         \includegraphics[scale=0.4]{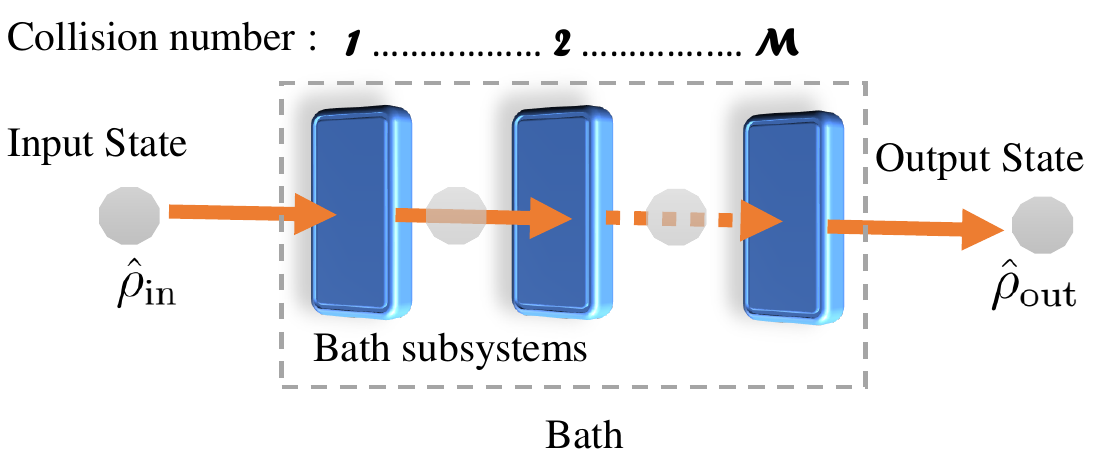}\\(a)
         \label{fig:coll}
         \includegraphics[scale=0.6]{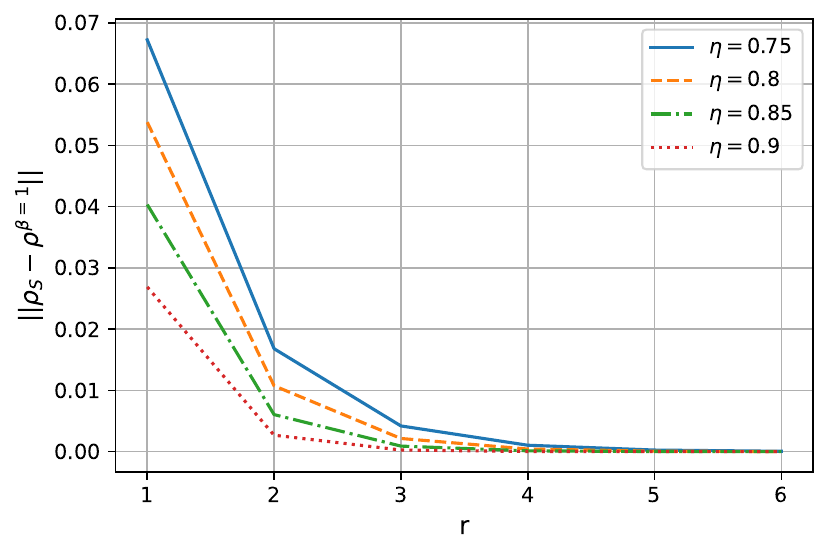}\\(b)
         \label{fig:GADtr}
         \caption{(a) Schematic diagram showing the collisional model with each step consisting of the GADC unitary ($U_{BS}^{\eta}$) acting on the probe and the bath, each of whose subsystems are in a Gibbs state at a fixed temperature. Equivalently, one could consider the unitary interaction $U_{BS}=U_{BS}^{\eta}\otimes I_A$ between one of the (purified) bath subsystems and the probe at each ``collision". (b) Trace distance between the state of the probe and a thermal state as a function of the collision number in the collisional GADC model. The probe is initialized in the state $|0\rangle$, whereas the initial state of each of the bath subsystems is defined as $|\theta^{\beta=1}\rangle$ corresponding to temperature T=1. $\Delta E =1$ for both probe and bath subsystems.}
        \label{fig:coll_scheme}
\end{figure}

In this model of gradual thermalisation, the state of a complete bath is of the form $|\theta^\beta\rangle^{\otimes\mathcal{M}}~,$ where $\mathcal{M}$ is the number of bath subsystems, and we have considered each of the bath subsystems to be in a purified state (for both one- and two-bath scenarios). We call an individual application of the unitary $U_{BS}$ (on the probe and purified bath subsystem) a collision and assign a ``collision number'' to each such operation, see Fig.~\eqref{fig:coll_scheme}.

Such modelling of the thermalisation process enables us to be in the pre-thermalisation regime by controlling the strength and/or the number of collisions. Fig.~\eqref{fig:coll_scheme}, shows the trace distance between the state of the probe, initialized as $|0\rangle$ (the ground state), and a thermal state at a temperature $T=1$, as a function of the number of collisions. This distance approaches zero as the number of collisions increase. The number of collisions after which the trace norm falls below a fixed threshold $||\rho_S-\rho^{\beta=1}||< \epsilon$, signalling that the probe approaches thermalisation, depends on both the interaction strength and the initial state of the probe. 

In the following subsection, we use the above model to illustrate the drop in visibility in the pre/post thermalisation regime and the key differences while focusing on the two scenarios identified in the previous sections---the one-bath and two-bath cases. In all instances shown here, we consider any local unitaries or isometries acting on the bath(s) to be the identity. This in particular implies that the bath does not have its own intrinsic dynamics and that the purifications considered in the single-bath case are likewise independent of the control.

\subsection{Two bath case}
In the two-bath case of Sec.~\ref{sec:TwoBath} the probe interacts with two separate bath subsystems---for example, localised in the arms of a MZ interferometer Fig.~\ref{fig:2BathMZInterferometer}. For the study of pre-thermalisation in this model it will be convenient to use purified states of the baths, by which the initial state of both baths is given as
\begin{equation}
|B\rangle = |\theta^{\beta_0}\rangle^{\otimes\mathcal{M}} \otimes |\theta^{\beta_1}\rangle^{\otimes\mathcal{M}}~,
\end{equation}
where each state $\ket{\theta^{\beta_i}}$ is defined as in Eq.~\eqref{eq:psitfd}.
 
The interaction between the probe, the control, and the baths takes the form of Eq.~\eqref{eq:fullU} where the probe-bath unitaries are now additionally split into a tensor product of terms acting at subsequent collisions. Hence, for the $r^{\mathrm{th}}$ collision we have:   
\begin{align}
U_{B_0S}^r &=(U_{BS})^{r}\otimes(\mathbb{I}_2\otimes\mathbb{I}_2)^{\otimes \bar{r}}~,\nonumber\\
U_{B_1S}^r &=(U_{BS})^{r}\otimes(\mathbb{I}_2\otimes\mathbb{I}_2)^{\otimes \bar{r}}~,
\label{eq:uni_2bath}
\end{align}
where $U_{BS}$ denotes the unitary operator in Eq.~\eqref{eq:USB}, superscript $r$ indicates the unitary being confined to the $r^{\mathrm{th}}$ subsystem, and $\bar{r}$ refers to the remaining subsystems of $\mathcal{M}$. The subscript 2 on the identity operator reminds us that we are modelling each of the $\mathcal{M}$-subsystems of either of the baths as a qubit.

\begin{figure}[t]
	\includegraphics[width=\columnwidth]{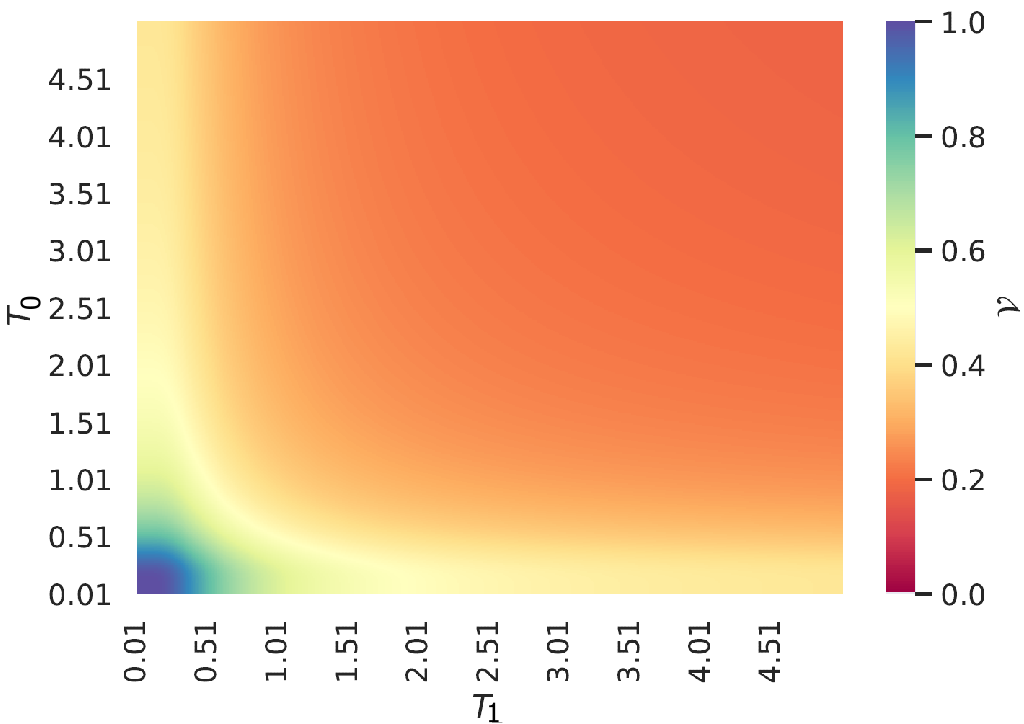}\\(a)
	\label{fig:2Bathpart_temp}
	\includegraphics[width=\columnwidth]{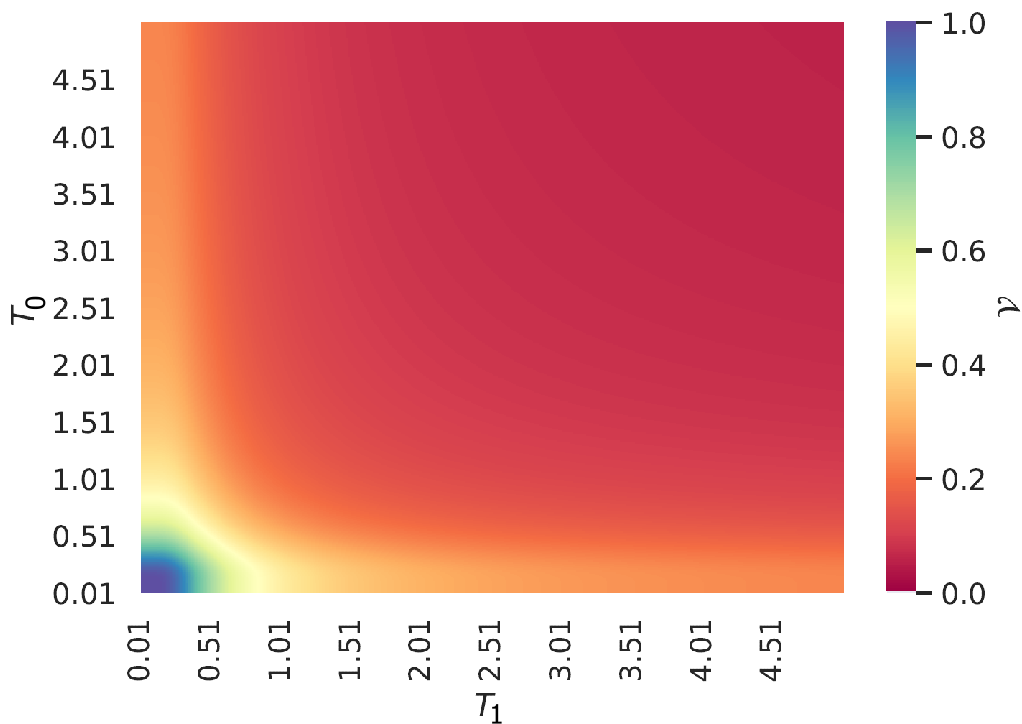}\\(b)
	\label{fig:2Bathfull_temp}
	\caption{Heat map of the visibility of the control as a function of the temperatures of the baths in the two-bath case for (a) pre-thermalisation, $\mathcal{M}=3$ and (b) post-thermalisation, $\mathcal{M}=5$. The interaction parameter is set: $\eta = 0.8$ to allow for partial thermalisation.}
	\label{fig:2_bath}
\end{figure}

We compare the pre- and post-thermalisation visibilities for the two-bath case (cf  Sec.~\ref{sec:TwoBath}) by changing the number of collisions (no.~of subsystems within the bath). The results are presented in Fig.~\ref{fig:2_bath}. The interaction strength is kept constant and we take the energy ground state as the probe's initial state. Notably, we see a very good agreement between full- and pre-thermalisation scenarios. In particular, we find that in the pre-thermalisation case, even if the temperatures of the two baths are equal, the visibility is significantly reduced, with a stronger reduction for higher temperatures. This is in contrast with the one-bath case in Fig.~\ref{fig:1_bath}, to be discussed in the next section, where for equal temperatures the visibility is maximal. Note also that the visibility generally falls with the number of collisions while the high-visibility parameter region shrinks. These results are in agreement with the analytical results for full thermalisation in Sec.~\ref{sec:TwoBath}.

\subsection{One bath case }
As outlined in Sec.~\ref{sec:OneBath}, in the one bath case the purified state of the bath is entangled with the control DoF, and its initial state is
\begin{equation}
\ket{B}= \ket{0}_C \ket{\theta^{\beta_0}}^{\otimes\mathcal{M}}+\ket{1}_C \ket{\theta^{\beta_1}}^{\otimes\mathcal{M}}~.
\label{eq:bath}
\end{equation}
Since here we assume that the purification of the bath is in the same basis, the operator $V^{01}_{bb'}$ is an identity. Note that in the $r^{\mathrm{th}}$ collision the probe interacts with the $r^{\mathrm{th}}$ subsystem of the bath and thus the unitary operator is of the form $(U_{BS})^{r}\otimes(\mathbb{I}_2\otimes\mathbb{I}_2)^{\otimes \bar{r}}$, as in Eq.~\eqref{eq:uni_2bath}.

\begin{figure}
	\includegraphics[width=\columnwidth]{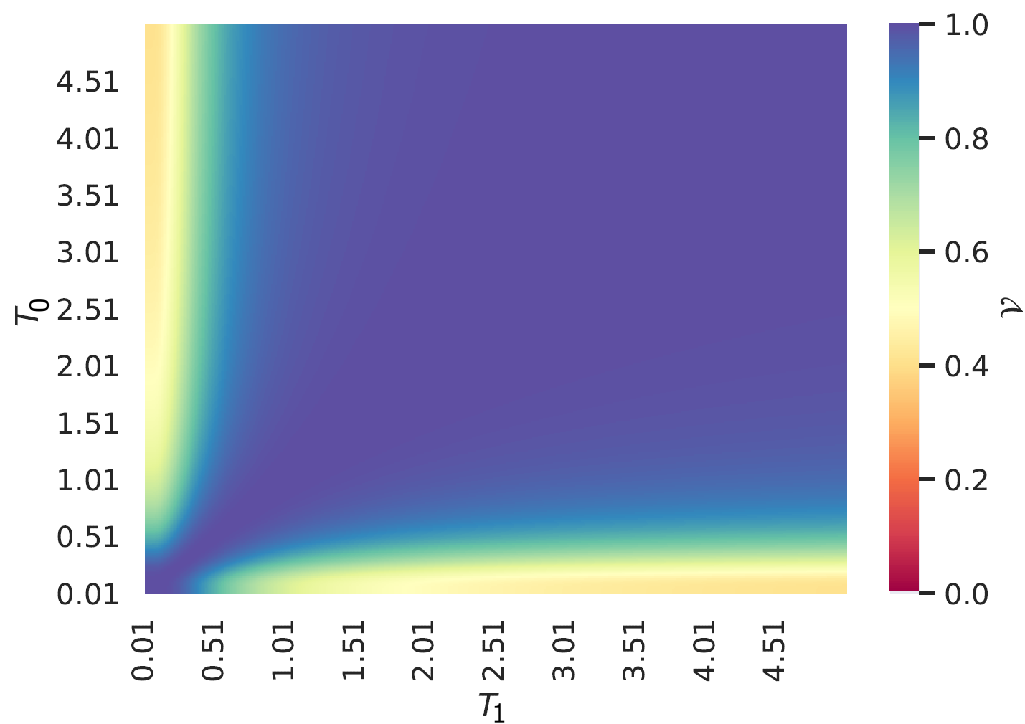}\\(a)
	\label{fig:1Bathpart_temp}
	\includegraphics[width=\columnwidth]{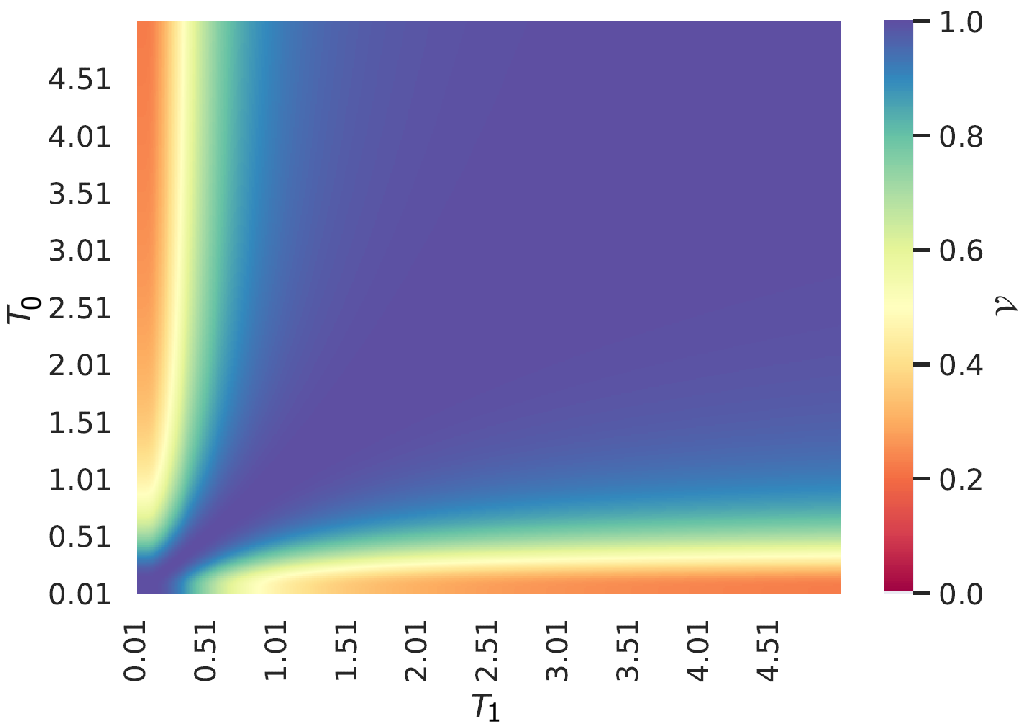}\\(b)
	\label{fig:1Bathfull_temp}
	\caption{Heat map of the visibility of the control as a function of temperatures of the bath in the one-bath case for (a) pre-thermalisation, $\mathcal{M}=3$ and (b) post-thermalisation, $\mathcal{M}=5$. The interaction parameter is set: $\eta = 0.8$ to allow for partial thermalisation.}
	\label{fig:1_bath}
\end{figure}

The comparison between the visibilities obtained for partially thermalising maps and completely thermalising maps for the one-bath case is shown in Fig.~\ref{fig:1_bath}. Once again, these results have been obtained by changing the number of collisions, keeping the interaction strength constant, and taking the probe in its energy ground state as the initial state. We find that the visibility is the greatest for similar temperatures of the baths and decreases when the temperatures are different. This decrease is steeper for higher numbers of collisions and therefore depends on the fact that the probe is not fully thermalised. Furthermore, the parameter space of $T_0$ and $T_1$, where the visibility is close to its maximum, shrinks as the number of collisions increases, signalling an increase in the distinguishability and the resulting loss of coherence.

\section{Returning to Relativity}\label{sec:Rel}
As discussed at the beginning of this work, there are certain highly relevant scenarios in relativistic quantum thermodynamics which naturally lead to the two cases we have studied here. For example, the two-bath model would arise in extended thermal systems in general relativity: Due to the Tolman-Ehrenfest effect~\cite{Tolman1930, TolmanEhrenfest1930}, a bath in thermal equilibrium in a gravitational field has a non-uniform temperature $T_i$ as measured by a local thermometer, where $T_i\sqrt{-g_{00}(r_i)}=\text{const}$. 
This is also true for extended quantum baths in equilibrium~\cite{Robles2017}. Therefore, a probe in a superposition of different locations would effectively interact with different baths at different temperatures, with the bath closest to the gravitational source experiencing greater time dilation, and therefore having higher temperature.

So if our two baths were at different heights in a gravitational field, e.g.~due to a Schwarzschild black hole, fixed observers at radii $r_0$ and $r_1$ would see temperatures:
\begin{equation}
T_0 = \frac{T_H}{\sqrt{-g_{00}(r_0)}}~;~~
T_1 = \frac{T_H}{\sqrt{-g_{00}(r_1)}},
\end{equation}
where in this case the generalised constant is the Hawking temperature $T_H$, and if $r_0 < r_1$, then $T_0 > T_1$. 
We expect the final state of the probe after interacting with such an environment to be similar to that of our two-bath case, Eqs~\eqref{eq:2bathfinaliso} or~\eqref{eq:2bathextraunitaries}.

The second model would in turn arise in relativistic quantum systems subject to acceleration. As mentioned in our introduction, internal DoFs of a system on a trajectory with constant proper acceleration experience the Unruh effect~\cite{Unruh1976, Crispino2008}. They thermalise to an acceleration-dependent temperature upon interaction with a quantum field, even though the field is in a vacuum state according to any inertial observer. To derive this prediction one can express a vacuum state in terms of the field modes in an accelerated reference frame (also called Rindler modes). The inertial vacuum is a two-mode squeezed state of these modes---a thermofield double state, or purification, as discussed in Section~\ref{sec:defineprob}. Rindler modes can be divided into ``left'' and ``right'' modes, in reference to the two sides of a ``Rindler horizon''\footnote{Such a horizon is formed by light-like trajectories to which an accelerating worldline asymptotes in the infinite future and infinite past.}. These modes depend on where the acceleration takes place in addition to the magnitude of acceleration since both criteria define the location of the horizon.

A superposition of trajectories, such as in Refs~\cite{Foo2020, Barbado2020, fooPRR2021}, leads to the decomposition of the vacuum state as: 
\begin{equation}\label{eq:trajsup}
\sum_{x=0,1} C \sum_{n=0}^{+\infty} e^{-\pi n \omega/a} | n, R(x) \rangle | n , L(x) \rangle |x\rangle,
\end{equation}
where $C = \sqrt{1- e^{-2\pi\omega/a}}$ is a normalisation factor, $a$ is the magnitude of the acceleration and $|n,R (L)\rangle$ is an $n$-particle state with frequency $\omega$ in the right (left) wedge. 
This corresponds to an Eq.~\eqref{eq:psitfd} purification, where the field modes constitute the bath and the bath's purified state is the above-mentioned vacuum of the relativistic field. The difference between the purifications is due to the difference between field modes associated with the Rindler wedges associated with different---e.g., rigidly translated---accelerated trajectories.
One can also see the clear correspondence between Eq.~\eqref{eq:trajsup} and the control-dependent ancillary system of Eq.~\eqref{eq:suppurificationPhi}, with $x=\{0,1\}$ in Eq.~\eqref{eq:trajsup} again the control---which also directly refers to the location of the accelerated trajectory---and the particle's internal state playing the role of the probe.  Importantly, the amplitude between the different purifications, e.g.~$\langle n_i , L(0) | n_j , L(1) \rangle $ in general depends on the overlap between the spacetime regions forming, in this example, the left Rindler wedge for the two different cases.   

\section{Discussion}\label{sec:discussion}
Motivated by the question of whether it is even possible to meaningfully characterise a ``superposition of temperatures'', we have explored here two ways to operationally make sense of such a notion. Both consist of some probe system undergoing thermalisation, where one model places the probe in a controlled superposition of interactions with either of two independent thermal baths, and the other model has the probe interacting with a single bath which is itself in a superposition of purifications of two thermal states.

We found that the two formulations lead to observably distinct behaviours, and so are not physically equivalent. Our results show that an extension of the notion of ``superposition'' beyond pure states requires a precise understanding of what physical scenario the superposition should capture. Specifically, we have found that the final state of systems undergoing thermalisation in superposition is not, in general, thermal. Furthermore, the thermalisation process depends on the physical implementation of the thermalisation channels (e.g., on the particular type of system-bath interaction). 
Technically, the sensitivity to the channel's implementation manifests as a dependence of observable properties (such as visibility) on the particular Kraus representation, or dilation, of the channel. This is a known feature in the general context of ``superpositions of quantum channels''---that is, in scenarios where channels are applied to subspaces, rather than systems or subsystems \cite{Oi2003, Araujo2014, Bisio2016, Chiribella2019, Abbott2020, Branciard2021}---but in the context of thermalisation only the consequences of applying two channels on the system in a control-state-dependent order have been previously explored.
Recent works have considered other thermodynamic effects in this so-called quantum switch scenario~\cite{Chiribella2013}, claiming that indefinite causal order does~\cite{Guha2020, Felce2020, Ban2021, Simonov2022}, or doesn't~\cite{Guerin2019, Abbott2020, Capela2023} offer advantages in engine/cycle performance, ergotropy, etc., compared to `ordinary' superpositions. See also ref.~\cite{Verma2024}.

Our results, beyond fundamental interest, show that experiments that probe thermal systems could be, by exploring quantum coherence as discussed in this work, sensitive to details of the thermal bath in ways ordinary thermometry can not be. 

Both of the scenarios we have considered feature a control system prepared in a superposition of two orthogonal states, each associated with one of the two temperatures. Intuitively, we think of different temperatures as distinct macroscopic variables, and one might expect that knowing the temperature could reveal enough information about the control to completely destroy its coherence. We have seen that this is generally not the case, and the precise prediction for the loss of coherence depends on the probe--bath interaction. Moreover, one can always find parameters for which the visibility vanishes, even for baths of equal temperatures, meaning that the which-way information leaking to the environment is not encoded in temperature alone. On the other hand, the \textit{maximum} visibility remains strictly below one when the two temperatures in superposition are different, giving a meaningful measure of how temperature reveals which-way information.

As discussed in the previous section, the models we have considered here could arise in certain relativistic quantum physics scenarios, e.g.~a quantum extension of the Tolman-Ehrenfest problem~\cite{TolmanEhrenfest1930}, or the Unruh and Hawking effects for particle detectors with quantised centre of mass~\cite{Foo2020, Barbado2020,fooPRR2021}. In fact, the one-bath model developed in this work explains previous results in which a local accelerating probe does not thermalise even if moving in a spatial superposition of trajectories with the exact same proper acceleration. As such trajectories are related by a rigid translation, and the global vacuum state of the field is translation-invariant, such a result did not have an intuitive explanation and seemed at odds with the underlying translational symmetry of the problem. The present work shows that this result simply arises from the fact that the probe effectively interacts with different subsystems of the relativistic field and so thermalisation does not necessarily follow from translation invariance.

Finally, as shown in Sec.~\ref{sec:partialpre}, we see our results do not depend on full thermalisation of the probe. Thus they will be relevant for physical, non-idealised scenarios, including those relativistic scenarios discussed.

\acknowledgements
This work was supported by Australian Research Council (ARC) Future Fellowship grant FT210100675, ARC DECRA grant DE170100712, and through the ARC Centre of Excellence for Engineered Quantum Systems (EQUS) CE170100009, as well as the Knut and Alice Wallenberg Foundation through Wallenberg Academy Fellowship No. 2021.0119. Nordita is supported in part by NordForsk. The authors acknowledge the traditional owners of the land on which the University of Queensland is situated, the Turrbal and Jagera people.

\appendix
\section{Two-bath Scenario with isometries as unitaries}\label{sec:appendix}

Recall the Kraus operator form of a regular quantum channel given in Eq.~\eqref{eq:thermalchannel} in the main text:
\begin{equation}
\text{Tr}_B \big\lbrace U_{B_iS} \rho_{BS} U_{B_jS}^{\dagger} \big\rbrace = \sum_{k} c_{k}^{\beta_i} M_{k k}^i \rho_S \sum_{l} c_{l}^{\beta_j} M_{l l}^{j\dagger}.
\end{equation} 

For the two-bath scenario of Section~\ref{sec:TwoBath}, the full final state of the probe in terms of the $M_{kl}=\bra{k}U_{BS}\ket{l}$ Kraus operators is then
\begin{align}
\begin{split}\label{eq:2bathfinal+Kraus}
\rho_S{(\phi)} = \frac{1}{4} \sum_{k,l} \Big(  c_{l}^{\beta_0} M_{k l}^0 \rho_S M_{k l}^{0\dagger} +  c_{l}^{\beta_1} M_{k l}^1 \rho_S M_{k l}^{1\dagger}\\
  e^{i\phi} c_{k}^{\beta_0} c_{l}^{\beta_1} M_{kk}^0 \rho_S   M_{l l}^{1\dagger} + e^{-i\phi} c_{k}^{\beta_1}c_{l}^{\beta_0} M_{kk}^1 \rho_S   M_{l l}^{0\dagger} \Big).
\end{split}
\end{align}

We now consider the isometries that relate different Kraus representations of a channel to be unitaries. We introduce arbitrary unitary matrices $u^i$, with matrix elements $u^i_{kl}$, such that the new Kraus operators are ${M'}^i_{kl}=\sum_s u^i_{ks}\ket{l}_S \bra{s}_S$, notably different for each of the two bath subsystems. This, along with the SWAP operator interaction discussed in the main text, yields a final state of the probe subject to a quantum superposition of thermal channels, 
\begin{equation}\label{eq:2bathextraunitaries}
\rho_S{(\phi)} = \frac{1}{4}\Big[\rho_S^{\beta_0} + \rho_S^{\beta_1} + \Big(e^{i\phi}\rho_S^{\beta_0}u^0\rho_Su^{1\dagger}\rho_S^{\beta_1} + \text{H.c.}\Big)\Big].
\end{equation}

As in the main text Section~\ref{sec:TwoBath}, even if both baths are at the same temperature, the system still does not thermalise, as explicitly seen from Eq.~\eqref{eq:2bathextraunitaries} for $\beta_0 = \beta_1 \equiv \beta$.

Indeed, for the final state to be thermal, we need $\rho_S^{\beta}u^1\rho_Su^{0\dagger}\rho_S^{\beta}=e^{i\alpha}\rho_S^{\beta}$, where $\alpha\in\mathbb{R}$ is a possible additional phase. Taking matrix elements in the energy eigenbasis, the above requires $e^{i\alpha}c_k^\beta\delta_{k,l}=c_k^\beta c_l^\beta \bra{k}u^1\rho_Su^{0\dagger} \ket{l}$, for all $k,l$. However, due to the non-negativity of the thermal weights $c_k^\beta$, state normalisation, and the unitarity of $u^i$, the modulus of the right-hand side is strictly less than $c_k^\beta$. That is to say, unless 1) $c_k^\beta=\delta_{k,0}$ (indicating both baths are at zero temperature), 2) the initial probe state $\rho_S$ is pure, and 3) the unitaries $u^i$ rotate the probe state to the energy ground state, and are thus equal (up to a global phase $\alpha$), the final state cannot be thermal.

As mentioned in the main text, the visibility is quantified by the magnitude of the off-diagonal elements of the final state, here of the control~\cite{Mandel1991, Englert1996}. Equivalently, it is the contrast of the interference pattern obtained by measuring the control in a superpostion basis. 
The probability of measuring the control in the state $\ket\phi$ is $\text{P}{(\phi)} = \text{Tr}\rho_S{(\phi)}$, and for our present case reads
\begin{equation}\label{eq:2bathprob}
\text{P}{(\phi)} = \frac{1}{2} + \frac{1}{2}  \left|\text{Tr} \left\{u^{0\dagger}\rho_S^{\beta_0} \rho_S^{\beta_1} u^1 \rho_S   \right\}\right|\cos(\phi+\psi),
\end{equation}
where $\psi$ is a phase defined via $\text{Tr}_S \{u^{0\dagger}\rho_S^{\beta_0} \rho_S^{\beta_1} u^1 \rho_S \}\equiv |\text{Tr}_S \{u^{0\dagger}\rho_S^{\beta_0} \rho_S^{\beta_1} u^1 \rho_S \}|e^{-i\psi}$. 
Hence, the visibility is 
\begin{equation}\label{eq:2bathvisibilityunitaries}
\mathcal{V} = \left|\text{Tr} \left\lbrace u^{0\dagger}\rho_S^{\beta_0} \rho_S^{\beta_1} u^1 \rho_S \right\rbrace \right|.
\end{equation}
Crucially, it depends not just on the temperatures of the two baths, but also the local unitaries. 

Note also that the visibility can never reach its maximum value of $1$, even if $\beta_0 = \beta_1$, except in the special case already discussed above, i.e.~where both baths are at zero temperature, the system is in a pure state and the unitaries $u^i$ rotate it to the ground state.

Moreover, for arbitrary fixed temperatures $\beta_i$ there exist $u^i$ which yield $\mathcal V=0$.
Indeed, decomposing the probe state in its eigenbasis $\rho_S =\sum_s p_s \ket{s}\bra{s}$, where $0\leq p_s\leq1$, and $\sum_s p_s=1$, Eq.~\eqref{eq:2bathvisibilityiso} reads
\begin{equation}\label{eq:2bathvisibilitykets}
\mathcal{V} = \left|\sum_s p_s\sum_{k} c^{\beta_0}_k c^{\beta_1}_k\bra{s} u^{0\dagger }\ket{k}\bra{k}u^1\ket{s} \right|.
\end{equation}
Taking $u^0$ to map each state $\ket{s}$ to some energy eigenstate, and $u^1$ to be $u^0$ times a cyclic permutation of energy eigenstates, means that if $\bra{k}u^0\ket{s}=1$ then $\bra{k}u^1\ket{s}=0$ and vice versa, resulting in $\mathcal{V} = 0$.

The above means that the minimum visibility over the different physical realisations of the thermalisation channel is zero. We have also seen that the coherence is in general not maximal. The question is therefore: what is the maximum of the visibility? To answer this question we find the maximum of $\mathcal{V}$ in Eq.~\eqref{eq:2bathvisibilitykets} over local bath unitaries $u^i$ for arbitrary but fixed temperatures of the baths and the probe initial state.

We let $\bra{s} u^{0\dagger}\ket{k} = \alpha^0_k(s)$ and $\bra{k}u^1\ket{s} = \alpha^1_k(s)$ and because $c_k^{\beta_i}$ as well as $p_s$ are non-negative, for maximum visibility, $\alpha^0_k(s)$ and $\alpha^1_k(s)$ must also be real and non-negative (up to a global phase). Hence Eq.~\eqref{eq:2bathvisibilitykets} becomes $\mathcal{V} = \sum_s p_s\sum_{k} c^{\beta_0}_k c^{\beta_1}_k\alpha^0_k(s)\alpha^1_k(s)$.

Starting with a pure state $\rho_S=\ket{s}\bra{s}$, we find the maximum visibility using the Lagrange multiplier method to find the optimum values of $\alpha^0_k$ and $\alpha^1_k$.

The Lagrangian $\mathcal{L} = \mathcal{\bar{V}}$ which we will optimise is
\begin{equation*}
\mathcal{\bar{V}} = \mathcal{V} - \lambda_0 \left(1- \sum_k \left|\alpha^0_k \right|^2\right) - \lambda_1 \left(1- \sum_k \left|\alpha^1_k \right|^2\right)
\end{equation*}
where $\sum_k\left|\alpha^0_k\right|^2 = 1$ and $\sum_k\left|\alpha^1_k\right|^2 = 1$.

With $d\bar{\mathcal{V}} = 0$ as a constraint, we arrive at a set of four simultaneous equations:

\begin{align}
\frac{d\bar{\mathcal{V}}}{d\alpha^0_l} &= c^{\beta_0}_l c^{\beta_1}_l \alpha^1_l + \lambda_0 2\alpha^0_l = 0\label{eq:LMalpha0} \\
\frac{d\bar{\mathcal{V}}}{d\alpha^1_l} &= c^{\beta_0}_l c^{\beta_1}_l \alpha^0_l + \lambda_1 2 \alpha^1_l = 0 \label{eq:LMalpha1}\\
\frac{d\bar{\mathcal{V}}}{d\lambda_0} &= -1 + \sum_k \left|\alpha^0_k \right|^2 = 0~~\implies \sum_k\left|\alpha^0_k\right|^2 = 1 \\
\frac{d\bar{\mathcal{V}}}{d\lambda_1} &= -1 + \sum_k \left|\alpha^1_k \right|^2 = 0~~\implies \sum_k\left|\alpha^1_k\right|^2 = 1
\end{align}

Working with equations \eqref{eq:LMalpha0} and \eqref{eq:LMalpha1},
\begin{align}
\alpha^0_k &= \frac{c^{\beta_0}_k c^{\beta_1}_k \alpha^1_k}{2 \lambda_0} \label{eq:alpha0}\\
\alpha^1_k &= \frac{c^{\beta_0}_k c^{\beta_1}_k \alpha^0_k}{2 \lambda_1} \label{eq:alpha1}
\end{align}

Substituting \eqref{eq:alpha1} into \eqref{eq:alpha0},
\begin{align*}
\alpha^0_k &= \frac{c^{\beta_0}_k c^{\beta_1}_k }{2 \lambda_0} \left(\frac{c^{\beta_0}_k c^{\beta_1}_k \alpha^0_k}{2 \lambda_1}\right) \\
&= \left(\frac{c^{\beta_0}_k c^{\beta_1}_k }{2}\right)^2 \frac{\alpha^0_k}{\lambda_0\lambda_1}
\end{align*}

This has the trivial solution $\alpha^0_k = 0$, which would imply $\alpha^1_k = 0$ too.

Otherwise, the above expression can also be rearranged for:
\begin{align*}
\lambda_0 \lambda_1 &= \left(\frac{c^{\beta_0}_l c^{\beta_1}_l}{2}\right)^2\\
\text{and} \qquad & \\
(\alpha^0_l)^2 &= \frac{\lambda_1}{\lambda_0}(\alpha^1_l)^2
\end{align*}

Which implies $\lambda_1 = \lambda_0$, which in turn implies
$\lambda_0 = \pm\frac{c^{\beta_0}_l c^{\beta_1}_l}{2}$. However, this means that $c^{\beta_0}_l c^{\beta_1}_l = \text{const}$, for all $l$. That is, $c^{\beta_0}_l = \text{const}$, $c^{\beta_1}_l = \text{const}$.

Since $\lambda_0 \lambda_1 = \left(\frac{c^{\beta_0}_l c^{\beta_1}_l}{2}\right)^2$ is the solution in the case of non-zero $\alpha_l^0$, we can only have one non-zero alpha as a solution.

With $\alpha_l^0 = \pm \alpha_l^1 = \pm \delta_{l,l'}$, then, $\mathcal{V}_{max} = c^{\beta_0}_{l'} c^{\beta_1}_{l'}$.

Since for a thermal state $c_l^{\beta_i} > c_{l+1}^{\beta_i}$, the maximum visibility is found to be
\begin{equation*}
\mathcal{V}_{\mathrm{max}}^{\mathrm{pure}} = c^{\beta_0}_0 c^{\beta_1}_0.
\end{equation*}

The intuition here, as mentioned in the main text, is that visibility is maximised when local unitaries effectively rotate the probe state $\ket{s}$ to the energy ground state (up to a phase), as it has the highest overlap with a thermal state at any finite temperature. 

For an arbitrary mixed state the maximum visibility then reads
\begin{equation}\label{eq:Vmax2bath}
\mathcal{V}_{max} =\sum_s p_s c^{\beta_0}_s c^{\beta_1}_s,
\end{equation}
where without loss of generality the probabilities $p_s$ are ordered decreasingly, $p_s\geq p_{s+1}$. This is because thermal weights are likewise ordered decreasingly, i.e.~$c_k^{\beta_i}>c_{k+1}^{\beta_i}$, and due to the orthogonality of the states $\ket s$ a unitary $u^i$ can only rotate one of the states from an orthogonal set to the energy ground state, but the same strategy can be iterated to find a unitary mapping of the eigenstates of $\rho_S$ to progressively higher energy eigenstates.

\providecommand{\href}[2]{#2}\begingroup\raggedright\endgroup


\begin{thebibliography}{10}

\bibitem{VinjanampathyAnders2016}
S.~Vinjanampathy and J.~Anders, ``Quantum thermodynamics,''
  \href{http://dx.doi.org/10.1080/00107514.2016.1201896}{{\em Contemporary
  Physics} {\bfseries 57}, 545--579 (2016)}.

\bibitem{Mehboudi2019}
M.~Mehboudi, A.~Sanpera, and L.~A. Correa, ``Thermometry in the quantum regime:
  recent theoretical progress,''
  \href{http://dx.doi.org/10.1088/1751-8121/ab2828}{{\em Journal of Physics A:
  Mathematical and Theoretical} {\bfseries 52}, 303001 (2019)}.

\bibitem{MillerAnders2018}
H.~J. Miller and J.~Anders, ``Energy-temperature uncertainty relation in
  quantum thermodynamics,''
  \href{http://dx.doi.org/https://doi.org/10.1038/s41467-018-04536-7}{{\em
  Nature Communications} {\bfseries 9}, 1--8 (2018)}.

\bibitem{Ghonge2018}
S.~Ghonge and D.~C. Vural, ``Temperature as a quantum observable,''
  \href{http://dx.doi.org/10.1088/1742-5468/aacfb8}{{\em Journal of Statistical
  Mechanics: Theory and Experiment} {\bfseries 2018}, 073102 (2018)}.

\bibitem{Stace2010}
T.~M. Stace, ``Quantum limits of thermometry,''
  \href{http://dx.doi.org/10.1103/PhysRevA.82.011611}{{\em Phys. Rev. A}
  {\bfseries 82}, 011611 (2010)}.

\bibitem{Jevtic2015}
S.~Jevtic, D.~Newman, T.~Rudolph, and T.~M. Stace, ``Single-qubit
  thermometry,'' \href{http://dx.doi.org/10.1103/PhysRevA.91.012331}{{\em Phys.
  Rev. A} {\bfseries 91}, 012331 (2015)}.

\bibitem{Tolman1930}
R.~C. Tolman, ``On the Weight of Heat and Thermal Equilibrium in General
  Relativity,'' \href{http://dx.doi.org/10.1103/PhysRev.35.904}{{\em Phys.
  Rev.} {\bfseries 35}, 904--924 (1930)}.

\bibitem{TolmanEhrenfest1930}
R.~C. Tolman and P.~Ehrenfest, ``Temperature Equilibrium in a Static
  Gravitational Field,'' \href{http://dx.doi.org/10.1103/PhysRev.36.1791}{{\em
  Phys. Rev.} {\bfseries 36}, 1791--1798 (1930)}.

\bibitem{Foo2020}
J.~Foo, S.~Onoe, and M.~Zych, ``Unruh-deWitt detectors in quantum
  superpositions of trajectories,''
  \href{http://dx.doi.org/10.1103/PhysRevD.102.085013}{{\em Phys. Rev. D}
  {\bfseries 102}, 085013 (2020)}.

\bibitem{Barbado2020}
L.~C. Barbado, E.~Castro-Ruiz, L.~Apadula, and {\v C}.~Brukner, ``Unruh effect
  for detectors in superposition of accelerations,''
  \href{http://dx.doi.org/10.1103/PhysRevD.102.045002}{{\em Phys. Rev. D}
  {\bfseries 102}, 045002 (2020)}.

\bibitem{fooPRR2021}
J.~Foo, S.~Onoe, R.~B. Mann, and M.~Zych, ``Thermality, causality, and the
  quantum-controlled Unruh--deWitt detector,''
  \href{http://dx.doi.org/10.1103/PhysRevResearch.3.043056}{{\em Phys. Rev.
  Research} {\bfseries 3}, 043056 (2021)}.

\bibitem{Lenard1978}
A.~Lenard, ``Thermodynamical proof of the Gibbs formula for elementary quantum
  systems,'' \href{http://dx.doi.org/10.1007/BF01011769}{{\em Journal of
  Statistical Physics} {\bfseries 19}, 575--586 (1978)}.

\bibitem{NielsenChuang2010}
M.~A. Nielsen and I.~L. Chuang, {\em Quantum computation and quantum
  information}.
\newblock Cambridge University Press, Cambridge ; New York, 10th anniversary
  ed.~ed., 2010.

\bibitem{KrausBook1983}
K.~Kraus, \href{http://dx.doi.org/10.1007/3-540-12732-1}{{\em States, effects
  and operations, vol. 190 of Lecture Notes in Physics}}.
\newblock Springer, Berlin, 1983.

\bibitem{Oi2003}
D.~K.~L. Oi, ``Interference of Quantum Channels,''
  \href{http://dx.doi.org/10.1103/PhysRevLett.91.067902}{{\em Phys. Rev. Lett.}
  {\bfseries 91}, 067902 (2003)}.

\bibitem{Abbott2020}
A.~A. Abbott, J.~Wechs, D.~Horsman, M.~Mhalla, and C.~Branciard,
  ``Communication through coherent control of quantum channels,''
  \href{http://dx.doi.org/10.22331/q-2020-09-24-333}{{\em {Quantum}} {\bfseries
  4}, 333 (2020)}.

\bibitem{Stinespring1955}
W.~F. Stinespring, ``Positive Functions on C*-Algebras,''
  \href{http://dx.doi.org/10.2307/2032342}{{\em Proceedings of the American
  Mathematical Society} {\bfseries 6}, 211--216 (1955)}.

\bibitem{TakahashiUmezawa1975}
Y.~Takahashi and H.~Umezawa, ``Higher order calculation in thermo field
  theory,'' {\em Collective phenomena} {\bfseries 2}, 55 (1975).

\bibitem{Israel1976}
W.~Israel, ``Thermo-field dynamics of black holes,''
  \href{http://dx.doi.org/https://doi.org/10.1016/0375-9601(76)90178-X}{{\em
  Physics Letters A} {\bfseries 57}, 107--110 (1976)}.

\bibitem{Crispino2008}
L.~C.~B. Crispino, A.~Higuchi, and G.~E.~A. Matsas, ``The {Unruh} effect and
  its applications,'' \href{http://dx.doi.org/10.1103/RevModPhys.80.787}{{\em
  Rev. Mod. Phys.} {\bfseries 80}, 787--838 (2008)}.

\bibitem{Mandel1991}
L.~Mandel, ``Coherence and indistinguishability,''
  \href{http://dx.doi.org/10.1364/OL.16.001882}{{\em Opt. Lett.} {\bfseries
  16}, 1882--1883 (1991)}.

\bibitem{Englert1996}
B.-G. Englert, ``Fringe Visibility and Which-Way Information: An Inequality,''
  \href{http://dx.doi.org/10.1103/PhysRevLett.77.2154}{{\em Phys. Rev. Lett.}
  {\bfseries 77}, 2154--2157 (1996)}.

\bibitem{Jeong2007}
H.~Jeong and T.~C. Ralph, ``Quantum superpositions and entanglement of thermal
  states at high temperatures and their applications to quantum-information
  processing,'' \href{http://dx.doi.org/10.1103/PhysRevA.76.042103}{{\em Phys.
  Rev. A} {\bfseries 76}, 042103 (2007)}.

\bibitem{Heinosaari2011}
T.~Heinosaari and M.~Ziman,
  \href{http://dx.doi.org/10.1017/CBO9781139031103}{{\em The Mathematical
  Language of Quantum Theory: From Uncertainty to Entanglement}}.
\newblock Cambridge University Press, 2011.

\bibitem{Rosati2018}
M.~Rosati, A.~Mari, and V.~Giovannetti, ``Narrow bounds for the quantum
  capacity of thermal attenuators,''
  \href{http://dx.doi.org/10.1038/s41467-018-06848-0}{{\em Nature
  Communications} {\bfseries 9}, 4339 (2018)}.

\bibitem{Scarani2002}
V.~Scarani, M.~Ziman, P.~\ifmmode \check{S}\else
  \v{S}\fi{}telmachovi\ifmmode~\check{c}\else \v{c}\fi{}, N.~Gisin, and
  V.~Bu\ifmmode~\check{z}\else \v{z}\fi{}ek, ``Thermalizing Quantum Machines:
  Dissipation and Entanglement,''
  \href{http://dx.doi.org/10.1103/PhysRevLett.88.097905}{{\em Phys. Rev. Lett.}
  {\bfseries 88}, 097905 (2002)}.

\bibitem{Robles2017}
S.~{Robles} and J.~{Rodr{\'\i}guez-Laguna}, ``{Local quantum thermometry using
  Unruh-DeWitt detectors},''
  \href{http://dx.doi.org/10.1088/1742-5468/aa60cd}{{\em Journal of Statistical
  Mechanics: Theory and Experiment} {\bfseries 3}, 033105 (2017)}.

\bibitem{Unruh1976}
W.~G. Unruh, ``Notes on black-hole evaporation,''
  \href{http://dx.doi.org/10.1103/PhysRevD.14.870}{{\em Phys. Rev. D}
  {\bfseries 14}, 870--892 (1976)}.

\bibitem{Araujo2014}
M.~Ara{\'{u}}jo, A.~Feix, F.~Costa, and {\v{C}}.~Brukner, ``Quantum circuits
  cannot control unknown operations,''
  \href{http://dx.doi.org/10.1088/1367-2630/16/9/093026}{{\em New J. Phys.}
  {\bfseries 16}, 093026 (2014)}.

\bibitem{Bisio2016}
A.~Bisio, M.~Dall'Arno, and P.~Perinotti, ``Quantum conditional operations,''
  \href{http://dx.doi.org/10.1103/PhysRevA.94.022340}{{\em Phys. Rev. A}
  {\bfseries 94}, 022340 (2016)}.

\bibitem{Chiribella2019}
G.~Chiribella and H.~Kristjánsson, ``Quantum Shannon theory with
  superpositions of trajectories,''
  \href{http://dx.doi.org/10.1098/rspa.2018.0903}{{\em Proc. R. Soc. A}
  {\bfseries 475}, 20180903 }.

\bibitem{Branciard2021}
C.~Branciard, A.~Cl\'{e}ment, M.~Mhalla, and S.~Perdrix,
  \href{http://dx.doi.org/10.4230/LIPIcs.MFCS.2021.22}{``{Coherent Control and
  Distinguishability of Quantum Channels via PBS-Diagrams},''} in {\em 46th
  International Symposium on Mathematical Foundations of Computer Science (MFCS
  2021)}, F.~Bonchi and S.~J. Puglisi, eds., pp.~22:1--22:20.
\newblock Schloss Dagstuhl -- Leibniz-Zentrum f{\"u}r Informatik, 2021.

\bibitem{Chiribella2013}
G.~Chiribella, G.~M. D'Ariano, P.~Perinotti, and B.~Valiron, ``Quantum
  computations without definite causal structure,''
  \href{http://dx.doi.org/10.1103/PhysRevA.88.022318}{{\em Phys. Rev. A}
  {\bfseries 88}, 022318 (2013)}.

\bibitem{Guha2020}
T.~Guha, M.~Alimuddin, and P.~Parashar, ``Thermodynamic advancement in the
  causally inseparable occurrence of thermal maps,''
  \href{http://dx.doi.org/10.1103/PhysRevA.102.032215}{{\em Phys. Rev. A}
  {\bfseries 102}, 032215 (2020)}.

\bibitem{Felce2020}
D.~Felce and V.~Vedral, ``Quantum Refrigeration with Indefinite Causal Order,''
  \href{http://dx.doi.org/10.1103/PhysRevLett.125.070603}{{\em Phys. Rev.
  Lett.} {\bfseries 125}, 070603 (2020)}.

\bibitem{Ban2021}
M.~Ban, ``Non-classicality created by quantum channels with indefinite causal
  order,''
  \href{http://dx.doi.org/https://doi.org/10.1016/j.physleta.2021.127381}{{\em
  Physics Letters A} {\bfseries 402}, 127381 (2021)}.

\bibitem{Simonov2022}
K.~Simonov, G.~Francica, G.~Guarnieri, and M.~Paternostro, ``Work extraction
  from coherently activated maps via quantum switch,''
  \href{http://dx.doi.org/10.1103/PhysRevA.105.032217}{{\em Phys. Rev. A}
  {\bfseries 105}, 032217 (2022)}.

\bibitem{Guerin2019}
P.~A. Gu\'erin, G.~Rubino, and i.~c.~v. Brukner, ``Communication through
  quantum-controlled noise,''
  \href{http://dx.doi.org/10.1103/PhysRevA.99.062317}{{\em Phys. Rev. A}
  {\bfseries 99}, 062317 (2019)}.

\bibitem{Capela2023}
M.~Capela, H.~Verma, F.~Costa, and L.~C. C\'eleri, ``Reassessing thermodynamic
  advantage from indefinite causal order,''
  \href{http://dx.doi.org/10.1103/PhysRevA.107.062208}{{\em Phys. Rev. A}
  {\bfseries 107}, 062208 (2023)}.

\bibitem{Verma2024}
H.~Verma and F.~Costa, ``Measuring two temperatures using a single
  thermometer,'' \href{http://arxiv.org/abs/2403.15186}{{\ttfamily
  arXiv:2403.15186 [quant-ph]}}.

\end{thebibliography}
\end{document}